\documentstyle[12pt,amsfonts]{article}
\parskip=\medskipamount

\newcommand {\cK}{{\cal K}}
\newcommand {\cL}{{\cal L}}

\newcommand {\cV}{{\cal V}}

\newcommand{\bQ}{{\bf Q}}

\newcommand{\bV}{{\bf V}}

\def\a{\alpha}
\def\b{\beta}

\def\d{\delta}
\def\e{\epsilon}

\def\G{\Gamma}

\def\l{\lambda}

\def\o{\omega}

\def\q{\theta}
\def\r{\rho}
\def\s{\sigma}

\def\z{\zeta}

\def\F{\Phi}
\def\J{\Psi}
\def\L{\Lambda}
\def\O{\Omega}

\def\S{\Sigma}
\def\U{\Upsilon}

\newcommand{\ad}{{\dot{\alpha}}}                           
\newcommand{\bd}{{\dot{\beta}}}                            
\newcommand{\ve}{\varepsilon}                            

\newcommand{\pa}{\partial}                           
\newcommand{\hf}{\frac12}

\newcommand{\sect}[1]{\setcounter{equation}{0}\section{#1}}

\newcommand{\be}{\begin{equation}}
\newcommand{\ee}{\end{equation}}
\newcommand{\bea}{\begin{eqnarray}}
\newcommand{\eea}{\end{eqnarray}}
\newcommand{\non}{\nonumber}
\begin{document}

\begin{titlepage}
\thispagestyle{empty}

\begin{flushright}
TP-TSU-9/98 \\
hep-th/9806147 \\
June, 1998
\end{flushright}

\vspace{1cm}
\begin{center}
{\Large\bf Projective Superspace as a Double--Punctured \\
Harmonic Superspace} \footnote{Presented at the Workshop on
Supersymmetries and Integrable Systems, Joint Institute for Nuclear
Research, Dubna, June 22--26, 1998.}\\
\end{center}

\begin{center}
{\bf Sergei M. Kuzenko}\\
\footnotesize{
{\it
Department of Physics,
Tomsk State University\\
Lenin Ave. 36, Tomsk 634050, Russia}\\
\tt{kuzenko@phys.tsu.ru}
}
\end{center}
\vspace{1cm}

\begin{abstract}
We analyse the relationship between the $N=2$ harmonic and
projective superspaces which are the only approaches
developed to describe general $N=2$ super Yang-Mills theories
in terms of off-shell supermultiplets with conventional
supersymmetry. The structure of low-energy
hypermultiplet effective action is briefly discussed.

\end{abstract}
\vfill

\end{titlepage}

\newpage
\setcounter{page}{1}

\sect{Introduction}

The $N=2$ harmonic superspace \cite{gikos} is the only manifesty
supersymmetric approach developed which allows us to describe general
$N=2$ super Yang-Mills theories and $N=2$ supergravity \cite{sugra}
in terms of unconstrained superfields. It is the harmonic superspace
which makes it possible to realize most general matter self-couplings
in $N=2$ supersymmetry \cite{gio} as well as to develop a general
setting for $N=2$ rigid supersymmetric field theories with gauged
central charge \cite{dikst}. Since all known $N=2$ supersymmetric
theories naturally emerge in the harmonic superspace approach,
this is a universal or `master' formalism for $N=2$ supersymmetry.

Harmonic superspace is also enough powerful
for the study of quantum aspects of $N=2$ super Yang-Mills theories.
The Feynman rules in harmonic superspace \cite{gios} have been
successfully applied
to compute the holomorphic corrections to the effective
action of $N=2$  Maxwell multiplet coupled to the matter
$q^+$ hypermultiplet \cite{bbiko} as well as the induced hypermultiplet
self-coupling \cite{ikz}. The background field formalism
in harmonic superspace \cite{bbko} has already been utilized to derive
Seiberg's holomorphic action for $N=2$ $SU(2)$
SYM theory \cite{seiberg} from $N=2$ supergraphs, to rigorously prove
the $N=2$ non-renormalization theorem \cite{bko} as well as to compute
the leading non-holomorphic quantum correction for $N=4$ $SU(2)$
SYM theory \cite{bkuz} \footnote{The non-holomorphic action for
$N=4$ $SU(2)$ SYM was first computed in \cite{pv,grr}.}.
On the other hand, the quantum harmonic formalism
has yet to be further elaborated. The main virtue
of this approach -- its universality --  often turns into
technical difficulties when making quantum loop calculations.
For example, to compute leading corrections to the effective action
we have to very carefully integrate out infinitely many auxiliary
degrees of freedom contained in the unconstrained analytic superfields
used to describe the charge hypermultiplet and $N=2$ vector multiplet,
and this is a non-trivial technical problem. But it seems that there exist
more economical techniques for computing the low-energy action,
which could be deduced from the first principles of the quantum harmonic
formalism.

Recently, a new manifestly $N=2$ supersymmetric approach
to quantum $N=2$ super Yang-Mills theories has intensively been
developed \cite{glrvw,gv,g,grr}. This approach is based on the concept of
$N=2$ projective superspace \cite{lr} having its origin in a remarkable
paper \cite{klr}.
A nice feature of the projective multiplets is that they can easily be
decomposed into a sum of well-known $N=1$ multiplets. Therefore,
the quantum calculations in $N=2$ projective superspace can be
controlled by comparing their results with those known for $N=1$
sypersymmetric models. On the other hand, it is far from obvious
how to reduce, say, the harmonic matter $q^+$ hypermultiplet
to $N=1$ superfields in an elegant way.
A drawback of the projective superspace
approach is that it allows us to manifestly realize only a $U(1)$
subgroup of the automorphism group $SU(2)_A$ of $N=2$ supersymmetry.
In the harmonic superspace approach, however, $SU(2)_A$ is
manifestly realized.

As far as we know, no detailed discussion has been
given in literature on the relationship between the harmonic and
projective superspaces. One can find a few comments in \cite{klr}
in the context of $N=2$ tensor multiplet models. It was also mentioned
\cite{lr} that the polar multiplet might be closely related to
the $q^+$ hypermultiplet as well as pointed out that
`the two approaches are presumably essentially equivalent'.
More recently, the two formalisms were
considered as alternative ones \cite{gv,g}.
In our opinion, these approaches are certainly
related and, in a sense, complementary to each other.
Therefore, it is worth combining their powerful properties for the study
of quantum $N=2$ supersymmetric field theories. The purpose of
the present paper is just to reveal such a relationship.

The paper is organized as follows. In Sections 2 and 3 we give a brief
introduction to projective and harmonic superspaces, respectively.
In Section 4 we suggest an approximation of the projective superfields
by smooth analytic superfields on harmonic superspace. Embedded into the
$q^+$ hypermultiplet, for instance, is a global analytic superfield
which coincides with the arctic multiplet inside a disk of radius $R$.
The pure arctic multiplet emerges in the limit $R \rightarrow  \infty$.
We derive projective actions from harmonic superspace and show, as an
example,
how to obtain the polar hypermultiplet propagator from that corresponding
to the $q^+$ hypermultiplet. The results of Section 4 imply in fact
that projective superspace provides us with a minimal truncation of
unconstrained analytic superfields, which inevitably breaks $SU(2)_A$ but
is nicely suited for $N=1$ reduction. In Section 5 we discuss the
projective truncation of a general low-energy $q^+$ hypermultiplet action
and show that the leading contributions to special truncated action
describe the so-called chiral--non-minimal nonlinear sigma-model
which was recently proposed to describe a supersymmetric low-energy
QCD action \cite{cnm}. In Appendix A we review the well-known
correspondence between tensor fields on $S^2$ and functions over
$SU(2)$ possessing definite $U(1)$-charges. Appendix B is devoted
to the harmonic superspace description \cite{gio} of $O(2k)$
multiplets.

\sect{Projective superspace}

Superfields living in the $N=2$ projective superspace \cite{lr}
are parametrized by a complex bosonic variable $w$ along with
the coordinates of $N=2$ global superspace ${\Bbb R}^{4|8}$
\be
z^M=(x^m,\theta_i^\alpha,
\bar \theta^i_{\dot{\alpha}}) \qquad \overline{\theta_i^\alpha}=\bar
\theta^{{\dot{\alpha}}\,i} \qquad i=1,2\;.
\ee
A superfield of the general form
\be
\Xi (z, w) = \sum_{n=-\infty}^{+\infty} \Xi_n (z) w^n
\label{generalprjsup}
\ee
is said to be projective if it satisfies the constraints
\be
\nabla_\a (w) \Xi (z, w) = 0 \qquad {\bar \nabla}_\ad (w)\Xi (z, w) = 0
\label{pc1}
\ee
which involve the operators
\be
\nabla_\a (w) \equiv  w D^1_\a - D^2_\a \qquad
{\bar \nabla}_\ad (w) \equiv {\bar D}_{\ad 1} + w {\bar D}_{\ad 2}
\label{nabla}
\ee
constructed from the $N=2$ covariant derivatives
$D_M = (\pa_m , D_\a^i , {\bar D}^\ad_i)$.
The operators $\nabla_\a (w)$
and ${\bar \nabla}_\ad (w)$ strictly anticommute with each other,
as a consequence of the covariant derivative algebra
\be
\{ D^i_\a , D^j_\b\} = \{ {\bar D}_{\ad i}, {\bar D}_{\bd j} \} = 0 \qquad
\{ D^i_\a , {\bar D}_{\ad j} \} = - 2 {\rm i} \, \d^i_j \, \pa_{\a \ad}\;.
\label{cda}
\ee
With respect to the inner complex variable $w$, the projective
superfields are holomorphic functions on the punctured complex plane
${\Bbb C}\,{}^*$
\be
\pa_{\bar w} \, \Xi (z, w) = 0\;.
\label{pholomorphy}
\ee

Constraints (\ref{pc1}) rewritten in components
\be
D^2_\a \Xi_n = D^1_\a \Xi_{n-1} \qquad
{\bar D}_{\ad 2} \Xi_n = - {\bar D}_{\ad 1} \Xi_{n+1}
\label{pc2}
\ee
determine the dependence of the component $N=2$ superfields
$\Xi$'s on $\q^\a_2$
and ${\bar \q}^2_\ad$ in terms of their dependence on $\q^\a_1$
and ${\bar \q}^1_\ad$. Therefore, the components $\Xi_n$ are
effectively superfields over the $N=1$ superspace parametrized by
\be
\q^\a = \q^\a_1 \quad \qquad {\bar \q}_\ad = {\bar \q}^1_\ad \;.
\ee
If the power series in (\ref{generalprjsup}) terminates somehow,
several $N=1$ superfields satisfy constraints involving
the $N=1$ covariant derivatives
\be
D_\a = D^1_\a \qquad \quad {\bar D}^\ad = {\bar D}^\ad_1 \;.
\ee

A natural operation of conjugation, which brings every projective
superfield into a projective one, reads as follows
\be
\breve{\Xi} (y,w) = \sum_{n} (-1)^n {\bar \Xi}_{-n} (z) w^{n}
\label{pconjugation}
\ee
with ${\bar \Xi}_n$ being the complex conjugate of $\Xi_n$.
A real projective superfield is constrained by
\be
\breve{\Xi} = \Xi \quad \Longleftrightarrow \quad
{\bar \Xi}_n = (-1)^n \Xi_{-n}\;.
\ee
The component $\Xi_0(z)$ is seen to be real.

Given a real projective superfield $\cL(z, w)$, $\breve{\cL} = \cL$,
we can construct a $N=2$
supersymmetric invariant by the following rule
\be
S = \int {\rm d}^4 x\, D^4 \cL_0 (z) | \qquad \quad
D^4 = \frac{1}{16} D^{\a 1} D^1_\a {\bar D}_{\ad 1}  {\bar D}^\ad_1
\ee
where $D^4$ is the $N=1$ superspace measure and
$U|$ means the $\q$-independent component of a superfield $U$.
Really, from the standard
supersymmetric transformation law
\be
\d \cL = {\rm i} \left(\ve^\a_i Q^i_\a +
{\bar \ve}^i_\ad {\bar Q}^\ad_i \right)  \cL
\ee
we get
\bea
\d S &=& {\rm i} \int {\rm d}^4 x\, \left(\ve^\a_i Q^i_\a +
{\bar \ve}^i_\ad {\bar Q}^\ad_i \right) D^4 \cL_0 | =
- \int {\rm d}^4 x\, \left(\ve^\a_i D^i_\a +
{\bar \ve}^i_\ad {\bar D}^\ad_i \right) D^4 \cL_0 | \non \\
&=& - \int {\rm d}^4 x\, \left(\ve^\a_2 D^2_\a +
{\bar \ve}^2_\ad {\bar D}^\ad_2 \right) D^4 \cL_0 | =
- \int {\rm d}^4 x\, D^4 \left(\ve^\a_2 D^2_\a +
{\bar \ve}^2_\ad {\bar D}^\ad_2 \right) \cL_0 |   \non \\
&=& - \int {\rm d}^4 x\, D^4 \left(\ve^\a_2 D^1_\a  \cL_{-1} |
- {\bar \ve}^2_\ad {\bar D}^\ad_1 \cL_1 | \right) = 0\;.
\non
\eea
The action can be rewritten in the form \cite{lr}
\be
S = \frac{1}{2\pi {\rm i}} \oint_{C} \frac{{\rm d}w}{w}
\int{\rm d}^4 x\, D^4 \cL |
\label{praction2}
\ee
where $C$ is a contour around the origin.

Let us review several multiplets which can be realized
in projective superspace \cite{lr}. It is worth starting with
the so-called polar multiplet (or $\U$ hypermultiplet) describing a charged
$N=2$ scalar multiplet
\footnote{An off-shell $N=2$ hypermultiplet is said to be
charged or complex if it possesses an internal $U(1)$ symmetry
that couples to complex Yang-Mills, and neutral or real otherwise;
neutral hypermultiplets can transform only in real representations of the
gauge group.}:
\be
\U (z,w) = \sum_{n=0}^{\infty} \U_n (z) w^n \qquad
\breve{\U} (z,w) = \sum_{n=0}^{\infty} (-1)^n {\bar \U}_n (z)
\frac{1}{w^n} \;.
\label{pm}
\ee
The projective superfields $\U$ and $\breve{\U}$ are called arctic
and antarctic \cite{glrvw}, repectively.
If we treat the components of $\U$ as $N=1$ superfields, then $\U_0$
is a chiral superfield, $\U_1$  a complex linear superfield,
and $\U_2, \U_3, \ldots,$ complex unconstrained superfields
\be
{\bar D}_\ad \U_0 = 0 \qquad  {\bar D}^2 \U_1 = 0 \;.
\ee
The corresponding super Lagrangian reads
\be
\cL = \breve{\U} \U \qquad
\cL_0 = \sum_{n=0}^{\infty} (-1)^n {\bar \U}_n \U_n\;.
\ee

Cutting off the power series in (\ref{pm}) at some finite stage $p>2$,
one results in the so-called complex $O(p)$ multiplet
\be
\L^{[p]} (z,w) = \sum_{n=0}^{p} \L_n (z) w^n \qquad
\breve{\L}^{[p]} (z,w) = \sum_{n=0}^{p} (-1)^n {\bar \L}_n (z)
\frac{1}{w^n} \;.
\label{co(k)}
\ee
Its component superfields are constrained as follows:
\bea
{\bar D}_\ad \L_0 = 0 & \qquad & {\bar D}^2 \L_1 = 0 \non \\
D_\a \L_p = 0 & \qquad & D^2 \L_{p-1} = 0
\eea
and the rest components are unconstrained. The case $p=1$ corresponds
to the on-shell hypermultiplet, while for $p=2$ we obtain two tensor
multiplets.

The next multiplet of principal interest is called \cite{glrvw} tropical
and looks as follows
\be
V (z,w) = \breve{V} (z,w) = \sum_{n=-\infty}^{+\infty} V_n (z) w^n \qquad
{\bar V}_n = (-1)^n V_{-n}
\label{tm}
\ee
with all the components being unconstrained $N=1$ superfields
but $V_0$ real. This projective superfield describes a free massless $N=2$
vector multiplet provided the corresponding gauge invariance is \cite{lr}
\be
\d V = {\rm i}\,( \breve{\S} - \S)
\label{tropgauge}
\ee
where $\S$ is an arbitrary arctic superfield. The gauge invariant action
reads \cite{g}
\be
S[V] = - \hf \int {\rm d}^{12} z \oint
\frac{{\rm d} w_1}{2\pi {\rm i}} \,\frac{{\rm d} w_2}{2\pi {\rm i}}\;
\frac{V(z, w_1)  V(z, w_2)} {(w_1 - w_2)^2}\;.
\label{pva}
\ee

Cutting off the power series in (\ref{tm}) at some finite stage $k>1$
but preserving the reality condition, one results in the so-called
real $O(2k)$
multiplet
\be
\O^{[2k]} (z,w) = \breve{\O}^{[2k]} (z,w) = \sum_{n=-k}^{+k} \O_n (z) w^n
\qquad  {\bar \O}_n = (-1)^n \O_{-n} \;.
\label{ro(k)}
\ee
The components are constrained by
\be
{\bar D}_\ad \O_{-k} = 0  \qquad  {\bar D}^2 \O_{-k+1} = 0 \qquad
{\bar \O_0} = \O_0
\ee
and $\O_{-n+2}, \ldots, \O_{-1}$ are unconstrained complex superfields.
The super Lagrangian
\be
\cL = \hf (-1)^k \left( \O^{[2k]} \right)^2  \qquad
\cL_0 =  \sum_{n=0}^{k} (-1)^{k-n} {\bar \O}_{-n} \O_{-n}
\ee
describes a real off-shell hypermultiplet. The case $k=1$, which
was excluded from our consideration, corresponds to the free $N=2$ tensor
multiplet.

Complex $O(p)$ multiplets (\ref{co(k)}) are of little importance
by themselves.
In the even case, $p=2k$, we can write
$\L^{[2k]} (z,w) = w^k \l^{[2k]} (z,w)$,
where $\l^{[2k]}$ is seen to be a complex combination of two
real $O(2k)$ hypermultiplets (\ref{ro(k)}). When $p$ is odd, on
the other hand, we cannot define a supersymmetric action with
the correct kinetic terms for all the chiral and complex linear superfields
contained in $\L^{[k]}$ \cite{glrvw}. However, the polar
or $O(\infty)$ multiplet is of principal importance, since it
provides us with a realization of charged hypermultiplet
and can be coupled to the $N=2$
gauge field in arbitrary representations of the gauge group.
A single charged hypermultiplet must inevitably possess
infinitely many auxiliary
fields \cite{hsw} in $N=2$ supersymmetry without central charge.

The $O(p)$ multiplets, $p>2$, have been intensively studied.
They were originally
formulated in the standard $N=2$ superspace \cite{klt}
(see also \cite{hsw})
in terms of symmetric
isotensors $\O^{(i_1 i_2 \cdots i_p)}(z)$ constrained by
\be
D_\a^{(i_1} \O^{i_2 \cdots i_{p+1})} =
{\bar D}_\ad^{(i_1} \O^{ i_2 \cdots i_{p+1})} = 0\;,
\ee
then described in harmonic superspace
\cite{gio} and finally realized in projective superspace \cite{lr}.
The $O(2k)$ multiplets provide us with different
off-shell realizations for real hypermultiplet.
Their harmonic superspace formulation \cite{gio} is briefly discussed
in Appendix B.

Since the harmonic and projective descriptions of the
$O(2k)$ multiplets are completely equivalent, in what follows
we will concentrate on answering to the question
whether there is a room for the polar and tropical multiplets
in the harmonic superspace approach.

\sect{Harmonic superspace}

Harmonic superspace ${\Bbb R}^{4|8} \times S^2$ is a homogeneous space
of the $N=2$ Poincar\'e supergroup. The most useful in practice
global parametrization of $S^2 = SU(2)/ U(1)$ is that in terms of
the harmonic variables ${u_i}^-\,,\,{u_i}^+$ which
parametrize $SU(2)$,
the automorphism group of $N=2$ supersymmetry,
\bea
& ({u_i}^-\,,\,{u_i}^+) \in SU(2)\non\\
& u^+_i = \ve_{ij}u^{+j} \qquad \overline{u^{+i}} = u^-_i
\qquad u^{+i}u_i^- = 1 \;.
\eea
As is demonstrated in Appendix A,
tensor fields over $S^2$ are in a one-to-one correspondence with
functions on $SU(2)$ possessing definite harmonic $U(1)$-charges. A function
$\J^{(p)}(u)$ is said to have the harmonic $U(1)$-charge $p$ if
$$
\J^{(p)}({\rm e}^{{\rm i}\varphi} u^+,{\rm e}^{-{\rm i}\varphi} u^-)=
{\rm e}^{{\rm i}\varphi p} \J^{(p)}(u^+,u^-) \qquad
|{\rm e}^{{\rm i}\varphi}| = 1\;.
$$
A function $\J^{(p)}(z,u)$ on ${\Bbb R}^{4|8}
\times S^2$ with $U(1)$-charge $p$ is called a harmonic $N=2$
superfield.

When working with harmonic superfields, it is advantageous to make use
of the operators
\bea
& D^{\pm \pm} =u^{\pm i} \pa / \pa u^{\mp i} \quad \qquad
D^0 = u^{+i} \pa / \pa u^{+i} - u^{-i} \pa / \pa u^{-i} \non
\\
& [D^0,D^{\pm \pm}] = \pm 2D^{\pm \pm} \qquad \quad [D^{++}, D^{--}] = D^0
\eea
being left-invariant vector fields on $SU(2)$. Here $D^{\pm \pm}$ are two
independent harmonic covariant derivatives on ${\rm S}^2$, while $D^0$ is
the U(1)-charge operator, $D^0 \F^{(p)} = p \F^{(p)}$.

Using the harmonics, one can convert the spinor covariant derivatives into
$SU(2)$-invariant operators on ${\Bbb R}^{4|8} \times S^2$
\be
D^\pm_\alpha = D^i_\alpha u^\pm_i
\qquad {\bar D}^\pm_{\dot\alpha}={\bar D}^i_{\dot\alpha} u^\pm_i
\;.
\ee
Then the covariant derivative algebra (\ref{cda})
implies the existence of the following
anticommuting subset $(D^+_\a, {\bar D}^+_\ad)$,
\be
\{D^+_\a,D^+_\b\} = \{ {\bar D}^+_ {\dot\a},
{\bar D}^+_{\dot\b}\}= \{D^+_\a, {\bar D}^+_{\dot\a} \}=0\;.
\label{zero}
\ee
As a consequence, one can define an
important subclass of harmonic superfields constrained by
\be
D^+_\a \F^{(p)}={\bar D}^+_{\dot\a} \F^{(p)}=0\;.
\label{anconstr}
\ee
Such superfields are functions over the so-called analytic
subspace of the harmonic superspace parameterized by
\be
\{\z, u^{\pm}_i \} \equiv\{
x^m_A,\q^{+\a},{\bar\q}^+_{\dot\a}, u^\pm_i \} \quad \qquad \F^{(p)}(z,u)
\equiv  \F^{(p)} (\z, u)
\label{analytsub}
\ee
where \cite{gikos}
\be
x^m_A = x^m - 2{\rm i} \q^{(i}\s^m {\bar \q}^{j)}u^+_i u^-_j \qquad
\q^\pm_\a = \q^i_\a  u^\pm_i
\qquad {\bar \q}^\pm_{\dot\a}={\bar
\q}^i_\ad u^\pm_i\;.
\label{analbasis}
\ee
That is why such superfields are called analytic.

The analytic subspace (\ref{analytsub}) is closed under
$N=2$ supersymmetry transformations
and real with respect to the generalized conjugation (called in \cite{bbiko}
the smile-conjugation)
$\; \breve{} \;\equiv \;\stackrel{\star}{\bar{}}$
\cite{gikos}, where the operation ${}^\star$ is defined by
$$
(u^+_i)^\star = u^-_i \qquad
(u^-_i)^\star = - u^+_i \quad \Rightarrow \quad
(u^{\pm}_i)^{\star \star} = - u^{\pm}_i
$$
whence
\be
(u^{+i}) \,\breve{{}} = - u^+_i  \qquad \quad (u^-_i)\,\breve{{}} = u^{-i}\;.
\label{breve}
\ee
The analytic superfields with even $U(1)$-charge
can therefore be chosen real.

Harmonic superspace provides us with the following universal, manifestly
$N=2$ supersymmetric action
\be
S = \int {\rm d}u \,{\rm d}\z^{(-4)}\, \cL^{(+4)} (\z,u) \qquad
\quad \breve{\cL}^{(+4)} = \cL^{(+4)}
\label{haction}
\ee
with $\cL^{(+4)}$ being a real analytic superfield of $U(1)$-charge $+4$.
Here the integration is carried out over the analytic subspace,
${\rm d}\z^{(-4)}={\rm d}^4x_A {\rm d}^2 \q^+ {\rm d}^2{\bar \q}^+$
and the integration over $SU(2)$ is defined by \cite{gikos}
\be
\int {\rm d}u \; 1 = 1 \qquad \int {\rm d}u \, u^+_{(i_1}
\ldots u^+_{i_n} u^-_{j_1}
\ldots u^-_{j_m)} = 0 \qquad n+m > 0\;.
\label{hint}
\ee

Let us review the three basic harmonic multiplets which are used
to realize general $N=2$ super Yang-Mills theories in terms of
unconstrained superfields \cite{gikos,gios}.
The $q^+$ hypermultiplet is formulated
in terms of an unconstrained analytic superfield $q^+ (\z,u)$
and its conjugate $\breve{q}^+ (\z,u)$ with the action
\be
S[q^+] = - \int {\rm d}u \,{\rm d}\z^{(-4)}\, \breve{q}^+ D^{++} q^+ \;.
\label{qaction}
\ee
The $q^+$ (charged) hypermultiplet can transform
in arbitrary representations of the gauge group. Using this multiplet,
one can constract most general matter self-couplings \cite{gio}.
Further, the $\o$ (real) hypermultiplet is formulated in terms of a real
unconstrained analytic superfield $\o (\z, u)$, $\breve{\o} = \o$,
with the free action
\be
S[\o ] = - \hf \int {\rm d}u \,{\rm d}\z^{(-4)}\, \left( D^{++} \o
\right)^2\;.
\label{oaction}
\ee
In eqs. (\ref{qaction}) and
(\ref{oaction}) the operator $D^{++}$ is to be chosen in the
analytic basis (\ref{analbasis}).
Finally, the free $N=2$ vector
multiplet is realized in terms of a real unconstrained analytic superfield
$V^{++}(\z,u)$, $\breve{V}^{++} = V^{++}$, endowed with the gauge
invariance
\be
\d V^{++} = - D^{++} \l
\label{hvm}
\ee
where $\l(\z,u)$ is an arbitrary real analytic scalar superfield.
The gauge invariant action reads
\be
S[V^{++}] = \hf \int {\rm d}^{12} z  {\rm d}u_1 {\rm d} u_2
\; \frac{V^{++}(z, u_1)  V^{++}(z, u_2)}{(u^+_1 u^+_2)^2} \;.
\label{hva}
\ee
The harmonic distributions such as $(u^+_1 u^+_2)^{-2}$ are defined
in \cite{gios}.

\sect{Embedding the projective superfields into \\ analytic superfields}

We turn to describing the precise relationship between the
projective and analytic superfields.

\subsection{Analytic superfields in local coordinates}
To start with, it is worth rewriting the properties of
analytic superfields in the local complex coordinates on $S^2$
introduced in Appendix A. Let $\F^{(p)}(z,u)$ be a smooth analytic
superfield with {\it non-negative} $U(1)$-charge $p$. In the north chart
it can be represented as follows
\be
\F^{(p)}(z,u) = (u^{+1})^p \; \F^{(p)}_{{\rm N}}(z,w, {\bar w})
\ee
where $\F^{(p)}_{{\rm N}}(z,w, {\bar w})$ is given as in eq. (\ref{fn}),
but now the corresponding Fourier coefficients
$\F^{(i_1 \cdots i_{n+p} j_1 \cdots j_n)} (z)$
are special $N=2$ superfields. The fact that $\F^{(p)}(z,u)$ is a
smooth function on ${\Bbb R}^{4|8} \times SU(2)$, is equivalent to
the requirement that
\be
\lim\limits_{|w| \to \infty} \; \frac{1}{w^p}
\F^{(p)}_{{\rm N}}(z,w, {\bar w})
\ee
is a smooth function on ${\Bbb R}^{4|8} $. Keeping in mind this boundary
condition, it is sufficient to work in the north chart only.

The operators $D^+_\a$ and ${\bar D}^+_\ad$ can be rewritten in the
manner \cite{klr}
\be
D^+_\a = - u^{+1} \,\nabla_\a (w) \quad \qquad
{\bar D}^+_\ad = - u^{+1} \, {\bar \nabla}_\ad (w)
\label{dtonabla}
\ee
where $\nabla_\a (w)$ and ${\bar \nabla}_\ad (w)$ are given by eq.
(\ref{nabla}). Therefore, the Grassmann analyticity requirements
(\ref{anconstr}) become
\be
\nabla_\a (w) \F^{(p)}_{{\rm N}}(z,w, {\bar w}) = 0 \quad \qquad
{\bar \nabla}_\ad (w)  \F^{(p)}_{{\rm N}}(z,w, {\bar w}) = 0\;.
\label{GA}
\ee
As is seen, the constraints do not specify the ${\bar w}$-dependence
of $\F^{(p)}_{{\rm N}}(z,w, {\bar w})$ at all. That is why we are in
a position to truncate analytic superfields in such a way
to result in projective superfields.

To represent the analytic action (\ref{haction}) in the local coordinates,
we first rewrite
\be
S = \int  {\rm d}^4 x {\rm d}u\, (D^-)^4\cL^{(+4)} (z,u)| \qquad
(D^-)^4 = \frac{1}{16}D^{-\a} D^-_\a {\bar D}^-_\ad {\bar D}^{-\ad}
\ee
and then notice
\be
(D^-)^4 \F^{(p)} = (\overline{u^{+1}})^4 \;\frac{(1+w{\bar w})^4}{w^2}\,
D^4 \F^{(p)}
\ee
for an arbitrary analytic superfield $\F^{(p)}$. In accordance with
eq. (\ref{ns}), in the overlap of the north and south charts
it is worth representing the real analytic Lagrangian $\cL^{(+4)} (z,u)$ as
\be
\cL^{(+4)} (z,u) = (u^{+1} u^{+2})^2 \,\cL^{(+4)}_{{\rm N-S}}
(z, w, {\bar w})
\ee
where $\cL^{(+4)}_{{\rm N-S}}$ is real with respect to the
smile-conjugation (\ref{defreal}). Finally, we notice the identity
\be
\int {\rm d}u \, f(u) = \frac{1}{\pi}
\int \frac{{\rm d}^2 w}{(1+w {\bar w})^2}\, f(w,{\bar w})
\ee
for any smooth function of vanishing $U(1)$ charge. As a result,
the action (\ref{haction}) turns into
\be
S =  \frac{1}{\pi}
\int  {\rm d}^4 x \int \frac{{\rm d}^2 w}{(1+w {\bar w})^2} \,
D^4\cL^{(+4)}_{{\rm N-S}} (z,w, {\bar w})| \;.
\label{haction2}
\ee

To complete our analysis, it is also necessary to express the operator
$D^{++}$ in the local coordinates on $S^2$. The importance of this
operator consists in the fact that $D^{++}$ moves every analytic
superfield into an analytic one. If $\F^{(p)} (u)$ is a function with
non-negative $U(1)$-charge $p$, in the north chart one readily gets
\be
D^{++} \F^{(p)} (u) = (u^{+1})^{p+2}\, (1+ w {\bar w})^2  \,
\pa_{\bar w} \F^{(p)}_{{\rm N}} (w, {\bar w})\;.
\label{dplus}
\ee

\subsection{From $q^+$ hypermultiplet to polar hypermultiplet}

Let us consider the equation
\be
D^{++} q^+ (\z, u) = 0\;.
\label{on-shell}
\ee
It defines the on-shell hypermultiplet provided $q^+$ is required to
be a global analytic superfield (a smooth superfield over the harmonic
superspace).
In this case the general solution to
(\ref{on-shell}) reads
\be
q^+ (z, u) = q^i (z)\, u^+_i \qquad \quad D^{(i}_\a q^{j)} =
{\bar D}^{(i}_\ad q^{j)} = 0\;.
\label{on-shell2}
\ee
But if we allow $q^+$ to be smooth everywhere
on $S^2$ but at the north pole, the general solution of (\ref{on-shell})
becomes
\be
{\bf q}^+ (z, u) = u^{+1} \U (z, w) \quad \qquad
\U (z, w) = \sum_{n=0}^{\infty} \U_n (z) w^n
\label{qsingular}
\ee
as a consequence of (\ref{dplus}). The analytic constraints
$D^+_\a {\bf q}^+ = {\bar D}^+_\ad {\bf q}^+ = 0$,
to which ${\bf q}^+$ is to be subjected,
tell us that $\U (z, w)$ is nothing else but the arctic multiplet
described in Section 2. Therefore, we obtain a local analytic
superfield being singular at the north pole of the two-sphere.

Let us introduce an isospinor $s^i$ and its conjugate ${\bar s}_i$
\bea
s^i = (1,0)  & \qquad \quad & s_i = (0,1) \non \\
{\bar s}_i = (1,0)   & \qquad \quad   & {\bar s}^i = (0,-1)
\label{punctures}
\eea
which corresponds to the south and north poles of $S^2$, respectively.
Then we can rewrite eq. (\ref{qsingular}) as follows
\be
{\bf q}^+ (z, u) = u^{+1} \sum_{n=0}^{\infty} \U_n (z)
\frac{(u^+s)^n}{(u^+ {\bar s})^n}
\ee
where $(u^+s) = u^{+i} s_i$ and completely similar for $(u^+ {\bar s})$.
In accordance with (\ref{breve}), the smile-conjugate of this superfield
reads
\be
\breve{{\bf q}}^+ (z, u) = u^{+2} \sum_{n=0}^{\infty} (-1)^n {\bar \U}_n (z)
\frac{(u^+{\bar s})^n}{(u^+ s)^n}
\ee
or, equivalently,
\be
\breve{{\bf q}}^+ (z, u) = u^{+2} \breve{\U} (z, w) \qquad \quad
\breve{\U} (z, w) = \sum_{n=0}^{\infty} (-1)^n {\bar \U}_n (z)
\frac{1}{w^n}\;.
\label{smileqsingular}
\ee
This superfield satisfies the constraints
$D^+_\a \breve{{\bf q}}^+ = {\bar D}^+_\ad \breve{{\bf q}}^+ = 0$,
since the smile-conjugation is analyticity-preserving.
Our consideration shows that ${\bf q}^+$ and $\breve{{\bf q}}^+ $
possess singularities at the north and south poles of $S^2$,
respectively. Therefore, this multiplet lives in
the two-point-punctured version ${\Bbb R}^{4|8} \times {\Bbb C}\,{}^*$
of harmonic superspace. It is also obvious that the superfield
$\breve{\U} (z, w)$ just introduced is exactly the antarctic
multiplet described in Section 2.

Since ${\bf q}^+$ and $\breve{{\bf q}}^+$ are holomorphic superfields,
the $q^+$ hypermultiplet action (\ref{qaction}) vanishes for such
superfields. But now we can construct another analytic Lagrangian
\be
L^{(+4)}[\U] = (u^+ {\bar s}) (u^+ s) \,\breve{{\bf q}}^+ {\bf q}^+
\ee
which is holomorphic on ${\Bbb C}\,{}^*$,
\be
D^{++} L^{(+4)}= 0\;.
\ee
It can be used to construct the following supersymmetric action
\be
S[\U] = \frac{1}{2\pi {\rm i}} \int \frac{{\rm d} w}{w}
\int {\rm d}\z^{(-4)} \,L^{(+4)}[\U]
\label{polaraction}
\ee
coinciding with the polar hypermultiplet action discussed in Section 2.

The polar hypermultiplet can be obtained as the limit of a sequence
of global analytic superfields. Let us introduce an auxiliary
smooth function $f_{R, \e}(x)$ of a real variable $x \in [0,  \infty)$:
\be
f_{R, \e}(x) = \left\{ \begin{array}{c c}
\exp \left( \frac{1}{x-R-\e} - \frac{1}{x-R} \right) \qquad \quad &
R < x < R + \e \\
0  \qquad \quad                       & x \in [0, R] \cup [R+\e,  \infty)
\end{array}     \right.
\ee
which we apply to construct another function
\be
F_{R, \e}(x) = \int_{x}^{R+\e} {\rm d}t \, f_{R, \e}(t) \Big/
\int_{R}^{R+\e} {\rm d}t \, f_{R, \e}(t)
\label{stepfunction}
\ee
to be used in what follows.
Here $R$ and $\e$ are `large' and `small' positive parameters,
respectively. The function $F_{R, \e}(x)$ is equal to one when
$0 \le x \le R$, decreases from one to zero when $R< x < R+\e$,
and is equal to zero when $x \geq R+\e$.

Now, we define global analytic superfileds $q^+_{R,\e}$ and
$\breve{q}^+_{R,\e}$ given in the north chart as follows:
\bea
q^+_{R,\e} (z, u) &=& {\bf q}^+ (z,u) \,F_{R,\e}(|w|) =
u^{+1}\, \U(z, w) \,F_{R,\e}(|w|) \non \\
\breve{q}^+_{R,\e} (z,u) &=& \breve{{\bf q}}^+(z,u) \,F_{R,\e}(|w|^{-1})
= u^{+2}\, \breve{\U}(z, w)
\,F_{R,\e}(|w|^{-1}) \;.
\label{apprpolar}
\eea
${}$For fixed parameters $R$ and $\e$, such superfields
form an {\it off-shell
multiplet} with respect to the $N=2$ supersymmetric transformations.
Let us assume that $R \gg 1$, $\e \ll 1$ and evaluate the $q^+$
hypermultiplet action (\ref{qaction}) for the superfields
(\ref{apprpolar}). Since $\U$ is holomorphic, the operator $D^{++}$
in (\ref{qaction}) acts on $F_{R,\e}(|w|)$ only. Accounting
the properties of $F_{R,\e}$, one observes that the integration over
$S^2$ produces a non-vanishing contribution only in a small region
enclosed between the two circles of radii $R$ and $R+\e$. If one
introduces real variables $\r$ and $\varphi$ defined by
$w = \r {\rm e}^{{\rm i} \varphi }$, the action can be brought to
the form
\be
S[q^+_{R,\e}] = - \frac{1}{2 \pi} \int {\rm d}^4 x \, D^4
 \int_{0}^{2\pi} {\rm d} \varphi
\int_{R}^{R+\e} {\rm d}\r \,
F_{R,\e}(\r^{-1}) \breve{\U}(z, w) \U(z, w)\, \pa_{\r}
F_{R,\e}(\r)\big| \;.
\ee
${}$From here one readily gets
\be
\lim\limits_{\e \to 0} \; S[q^+_{R,\e}] = S[\U]=
\frac{1}{2 \pi{\rm i}} \oint \frac{{\rm d}  w}{w}
\int {\rm d}^4 x \, D^4 \big(\breve{\U} \U \big) \big|\;.
\ee
Therefore, the polar multiplet action has its origin in
harmonic superspace.

\subsection{Projective action rule}
It is easy to derive the projective action rule (\ref{praction2})
from harmonic superspace. First of all, one should define
a global analytic real superfield $\cL^{(+4)}_{R,\e} (z,u)$ of $U(1)$-charge
$+4$, which looks like
\bea
\cL^{(+4)}_{R,\e} (z,u) &=& ( u^{+1} u^{+2} )^2\;
F_{R,\e}\big(|w|^{-1}\big) \, \cL(z, w) \,
F_{R,\e}\big(|w| \big) \non \\
&\equiv &( u^{+1} u^{+2} )^2\;
\cL_{R,\e}(z, w,{\bar w})\;.
\label{prlagapprox}
\eea
Associated with the analytic Lagrangian $\cL^{(+4)}_{R,\e}$ is the
supersymmetric action
\bea
S_{R,\e} &=& \int {\rm d}u \,{\rm d}\z^{(-4)}
\cL^{(+4)}_{R,\e} (\z,u) \non \\
  &=& \frac{1}{\pi}
\int  {\rm d}^4 x \int \frac{{\rm d}^2 w}{(1+w {\bar w})^2} \,
D^4\cL_{R,\e} (z,w, {\bar w})| \;.
\label{ptactionappr}
\eea
We then can represent
\be
\frac{1}{(1+w {\bar w})^2} \,D^4\cL_{R,\e} (z,w, {\bar w})|
= -\left(\pa_{{\bar w}} \frac{1}{(1+w {\bar w})} \right)\,
\frac{1}{w} \,D^4\cL_{R,\e} (z,w, {\bar w})|\;.
\ee
${}$Finally, it remains to make the following steps:
(i) integrate by parts in (\ref{ptactionappr}) (this is possible
since the function $w^{-1} \,D^4\cL_{R,\e} (z,w, {\bar w})|$ is regular
at $w=0$); (ii) account that $\cL(z,w)$ is holomorphic;
(iii) introduce the real variables $\r$ and $\varphi$ defined by
$w = \r {\rm e}^{{\rm i} \varphi }$. As a result, one observes
\be
\lim\limits_{\e \to 0} \;
\int {\rm d}u \,{\rm d}\z^{(-4)} \cL^{(+4)}_{R,\e} (\z,u) =
 \frac{R^2 - 1}{R^2 +1 } \;
\frac{1}{2\pi {\rm i}} \oint \frac{{\rm d} w}{w}
\int{\rm d}^4 x\, D^4 \cL |\;.
\ee

\subsection{Hypermultiplet propagators}

We have seen that the $q^+$ hypermultiplet and the $\U$ hypermultiplet are
closely related to each other. Therefore, there should exist a relationship
between their propagatots.

The $q^+$ hypermultiplet propagator \cite{gios} reads
\be
<q^+(z_1, u_1) \; \breve{q}^+ (z_2, u_2)> \;=\;
{\rm i} \;
\frac{(D^+_1)^4  (D^+_2)^4}{\Box} \;
\frac{\d^{12}(z_1 - z_2)}{(u^+_1 u^+_2)^3}
\label{qpropagator}
\ee
where
\be
(D^+)^4 = \frac{1}{16} D^{+\a} D^+_\a {\bar D}^+_\ad {\bar D}^{+ \ad}
\ee
and
\be
(u^+_1 u^+_2) = u_1{}^{+ i} {u_2}^+_i \;.
\ee
To compare the above propagator with that of the $\U$ hypermultiplet,
we are to express (\ref{qpropagator}) in the local coordinates on $S^2$.
For this purpose we represent
\bea
q^+(z, u) &=& u^{+1} \;q^+_{\rm N} (z, w, {\bar w}) \non \\
\breve{q}^+(z, u) &=& u^{+2}\; \breve{q}^+_{\rm S}
(z, y(w), {\bar y} ({\bar w}))
\equiv w \,u^{+1} \,\breve{q}^+_{\rm S} (z, w, {\bar w}) \;.
\eea
Further, we have to express the operators $D^+_\a$ and ${\bar D}^+_\ad$
via $\nabla_\a$ and ${\bar \nabla}_\ad$ by the rule (\ref{dtonabla}).
Finally, we should make use of the identity
\be
(u^+_1 u^+_2) = u_1{}^{+1}  u_2{}^{+1}
(w_1 - w_2) \;.
\ee
Therefore, we result with
\be
<q^+_{\rm N} (z_1, w_1, {\bar w}_1)\;
\breve{q}^+_{\rm S} (z_2, w_2, {\bar w}_2)> \;=\;
{\rm i} \;
\frac{(\nabla_1)^4  (\nabla_2)^4}{\Box} \;
\frac{\d^{12}(z_1 - z_2)}
{w_2\,(w_1 - w_2)^3} \;.
\ee
This expression coincides in form with the polar hypermultiplet
propagator \cite{glrvw}. Of course, the distributions $(w_1 - w_2)^{-3}$
which enter the $q^+$ and $\U$ hypermultiplet propagators are
defined on different functional spaces. But since we know how
to truncate the $q^+$ hypermultiplet to result with the
$\U$ hypermultiplet, one can immediately reproduce the
$\U$ hypermultiplet propagator without tedious calculations.

The above consideration was restricted to the case of massless
hypermultiplet, but it can be readily generalized to the massive case.
A general feature of $N=2$ off-shell hypermultiplets is that the
presence of a non-vanishing mass is equivalent to the coupling
to a background $N=2$ $U(1)$ vector multiplet with constant strength
\cite{gio,bbiko,ikz,buchkuz}. It is the mechanism which was
used for constructing the massive hypermultiplet propagators in
harmonic superspace \cite{ikz,buchkuz} and projective superspace
\cite{gv}.  Similarly to the massless case, the massive propagators
coincide in form.

\subsection{From $V^{++}$ multiplet to tropical multiplet}
Let us consider the equation
\be
D^{++} \cV^{++} = 0\;.
\ee
It describes the free $N=2$ tensor multiplet provided $\cV^{++}$
is required to be an analytic real superfield,
\be
D^+_\a \cV^{++} = {\bar D}^+_\ad \cV^{++} = 0 \qquad \breve{\cV}^{++}
= \cV^{++}\,,
\label{analandreal}
\ee
globally defined on harmonic superspace. In this case $\cV^{++}$
looks like
\be
\cV^{++}(z,u) = \cV^{(ij)}(z) u^+_i u^+_j \qquad
D_\a^{(i} \cV^{jk)} = {\bar D}_\ad^{(i} \cV^{jk)}=0
\ee
with $\cV^{(ij)}(z)$ being a real isovector superfield.
However, if one allows $\cV^{++}$
to be singular at the north and south poles, but keeps intact
the basic constraints (\ref{analandreal}), the general solution
becomes
\be
\bV^{++}(z, u) = {\rm i} u^{+1} u^{+2} V(z, w)
\label{trop2}
\ee
where $V(z,w)$ is now the tropical multiplet described in Section 2.
Because of the reality condition, $\bV^{++}$ cannot be singular only
at a single point.

The tropical multiplet is closely related to the analytic gauge superfield
which we briefly discussed in Section 3. To establish such a relationship,
let us introduce a special global analytic superfield $V^{++}_{R,\e}$
defined with help of the infinitely differentiable function
(\ref{stepfunction}):
\be
V^{++}_{R,\e} (z,u) =  F_{R,\e}\big(|w|^{-1}\big) \,\bV^{++}(z, u)
F_{R,\e}\big(|w| \big)
\;.
\label{vapprox}
\ee
In the limit $R \rightarrow \infty$, $V^{++}_{R,\e}$ turns into
$\bV^{++}$ (\ref{trop2}) defining the tropical multiplet.
The tropical action (\ref{pva}) can be derived from that
corresponding to the  analytic gauge superfield (\ref{hva}).
It is a simple exercise to prove the relation
\be
\lim\limits_{\e \to 0} \; S[V^{++}_{R,\e}] =
\left( \frac{R^2 - 1}{R^2+1} \right)^2 \;S[V] \;.
\ee
We see that the tropical multiplet action (\ref{pva}) has its
origin in the harmonic superspace approach.
To derive the tropical gauge transformation
(\ref{tropgauge}) from that corresponding to
the analytic gauge superfield (\ref{hvm}), let us choose $V^{++}_{R,\e}$
in the role of $V^{++}$ and
consider the following variation
\bea
\d V^{++}_{R,\e} & \equiv &
F_{R,\e}\big(|w|^{-1} \big) \, \d \bV^{++} \, F_{R,\e}\big(|w|\big) =
- F_{R,\e}\big(|w|^{-1} \big) \,
\big( D^{++} \L \big)\, F_{R,\e}\big(|w|\big)  \non \\
&=& - D^{++} \Big\{ F_{R,\e}\big(|w|^{-1} \big) \,
\L \, F_{R,\e}\big(|w|\big) \Big\}
+ \L D^{++} \Big\{ F_{R,\e}\big(|w|^{-1} \big) \,
F_{R,\e}\big(|w|\big) \Big\}
\label{modvar}
\eea
with the parameter $\L$ being defined as follows
\be
\L = \breve{\L} = \big(\hf + u^{+1} u^{-2} \big) \,
\big( \breve{\S} - \S \big)
\label{howsigmacomes}
\ee
where $\S(z, w)$ is required, for a moment, to be a projective
superfield only.
In the limit $R \rightarrow \infty$,
the variation (\ref{modvar}) formally turns into
the transformation law (\ref{tropgauge}). The fact that $\S$ must
be arctic is quite understandable.
As is obvious, the first term in the second line of (\ref{modvar})
does not contribute to the corresponding variation of $S[V^{++}_{R,\e}]$.
Keeping in mind this observation, one then finds
\be
\lim\limits_{\e \to 0} \; \d S[V^{++}_{R,\e}] =
\left( \frac{R^2 - 1}{R^2+1} \right)^2 \; \d S[V] \;.
\ee
Here $\d S[V]$ is the variation of the tropical multiplet action
with respect to  (\ref{tropgauge}).
The variation $\d S[V]$ vanishes only if $\S$ is an arctic superfield
\cite{g}.

Eq. (\ref{pva}) defines the linearized action of the pure
$N=2$ super Yang-Mills theory in the projective superspace approach.
By now, the full nonlinear action has not derived in terms of
the tropical prepotential $V(z,w)$. In principle, it can be
deduced from the well-known harmonic action for the
$N=2$ super Yang-Mills theory \cite{zupnik}, but with use of a
more delicate truncation than that considered above.

\subsection{Hypermultiplet coupled to abelian vector multiplet}
It is interesting to compare the harmonic and projective
off-shell realizations for a charged massless hypermultiplet
coupled to an abelian vector multiplet. In the harmonic superspace
approach, the action reads \cite{gikos}
\be
S[q^+, V^{++}] = - \int {\rm d}u \,{\rm d}\z^{(-4)}\,
\breve{q}^+ \left( D^{++} + {\rm i} V^{++} \right) q^+
\label{q-vaction}
\ee
and is invariant under the gauge transformations
\be
\d q^+ = {\rm i} \l \, q^+ \qquad \quad \d V^{++} = - D^{++} \l
\label{q-vtl}
\ee
with an arbitrary real scalar analytic parameter $\l$.
In the projective superspace approach, the action reads \cite{lr}
\be
S[\U, V]=
\frac{1}{2 \pi{\rm i}} \oint \frac{{\rm d}  w}{w}
\int {\rm d}^4 x \, D^4 \big(\breve{\U} {\rm e}^V \U \big) \big|
\label{u-vaction}
\ee
and the corresponding gauge invariance is
\be
\d \U = {\rm i} \S \, \U \qquad \quad
\d V = {\rm i}\,( \breve{\S} - \S)
\ee
where the gauge parameter $\S$ is an arctic superfield.

To link the two descriptions, it is worth replacing $\U$ by a local
analytic superfield
\be
\bQ^+ (z,u)= u^{+1} \exp \left( u^{-1} u^{+2} V(z,w) \right) \U(z,w)
\ee
which, in contrast to ${\bf q}^+$ (\ref{qsingular}),
possesses singularities at both poles
and is covariantly holomorphic,
\be
\left(D^{++} + {\rm i} \bV^{++} \right) \bQ^+ = 0
\ee
with $\bV^{++}$ defined as in eq. (\ref{trop2}).
The matter ($\bQ^+$) and gauge ($\bV^{++}$) superfields transform
similar to eq. (\ref{q-vtl}),
\bea
& \d \bQ^+ = {\rm i} \L \, \bQ^+
\qquad \quad \d \bV^{++} = - D^{++} \L \non \\
& \L = u^{-1} u^{+2} \breve{\S} - u^{+1} u^{-2} \S
\eea
but now the gauge parameter $\L$ becomes singular at the north and
south poles. To reproduce the action (\ref{u-vaction}), it is
sufficient to evaluate $S[\bQ^+_{R,\e}, \bV^{++}]$ in the limit
$\e \rightarrow 0$, where
\be
\bQ^+_{R,\e} (z,u) = \exp \left( u^{-1} u^{+2} V(z,w) \right)
q^+_{R,\e} (z,u)
\ee
with $q^+_{R,\e}$ defined as in (\ref{apprpolar}). Both $\bQ^+_{R,\e} $
and $\bV^{++}$ are not globally defined on the harmonic superspace.
However, the (manifestly gauge invariant) action $S[\bQ^+_{R,\e}, \bV^{++}]$
appears to be well defined, since either $\bQ^+_{R,\e} $ or its
conjugate $\breve{\bQ}^+_{R,\e}$ vanishes just in a small region where
$\bV^{++}$ and the other matter superfield become singular.

\sect{Low-energy hypermultiplet action}

In quantum $N=2$ super Yang-Mills theories, the effective dynamics
of hypermultiplets are described by a low-energy action of the general
form \cite{gio}
\be
S_{{\rm eff}} [q^+] = \int {\rm d}u \,{\rm d}\z^{(-4)} \,
\cK^{(+4)}_{{\rm eff}} \big(q^+, \breve{q}^+,
D^{++} q^+, D^{++} \breve{q}^+, \cdots, u \big) \;.
\ee
Under reasonable assumptions on the structure of
$\cK^{(+4)}_{{\rm eff}}$, this action can be readily truncated to
projective superspace to result with \cite{lr}
\be
S_{{\rm eff}} [\U] = \frac{1}{2\pi {\rm i}}
\oint \frac{{\rm d}w}{w} \int {\rm d}^4 x \, D^4 \cK_{{\rm eff}}
\big( \U, \breve{\U}, w \big) \big|\;.
\ee
In the simplest case when $\cK_{{\rm eff}}$ is $w$-independent,
one can immediately evaluate leading contributions to the low-energy
action which come from the physical $N=1$ chiral ($\F^I$) and complex
linear ($\G^I$) superfields contained in $\U^I$,
\bea
&\U^I(w) = \F^I + w \G^I \;\;+\;\; \mbox{auxiliary superfields} \non \\
&{\bar D}_\ad \F^I = 0 \qquad \quad {\bar D}^2 \G^I = 0\;.
\eea
One gets
\bea
S_{{\rm eff}} [\U] &=&   \frac{1}{2\pi {\rm i}}
\oint \frac{{\rm d}w}{w} \int {\rm d}^4 x \, D^4 \cK_{{\rm eff}}
\big( \U, \breve{\U} \big) \big|    \non \\
&=& \int {\rm d}^8 z \, \left\{ \cK_{{\rm eff}}
\big( \F, \bar{\F} \big) -  \G^I {\bar \G}^J \,
\frac{\pa^2}{\pa \F^I \pa {\bar \F}^J}
\cK_{{\rm eff}} \big( \F, \bar{\F} \big) \right\} + \ldots
\label{kahler}
\eea
This manifestly $N=2$ supersymmetric sigma-model possesses a K\"ahler
invariance of the form
\be
\cK_{{\rm eff}} \big( \U, \breve{\U} \big) \quad \longrightarrow \quad
\cK_{{\rm eff}} \big( \U, \breve{\U} \big) + \L \big( \U\big) +
{\bar \L} \big(\breve{\U} \big)
\ee
with an arbitrary holomorphic function $\L$. Hence $\cK_{{\rm eff}}
\big( \F, \bar{\F} \big)$ is a K\"ahler potential.
As is well-known, the target spaces of $N=2$ supersymmetric sigma-models
are hyper-K\"ahler manifolds \cite{agf}.
The case under consideration turns out to be very specific.
The physical scalars
$\F^I$ parametrize
a K\"ahler manifold, while $\G^I$ the tangent space at
the point $\{ \F^I \}$ of the K\"ahler manifold.
This follows from the fact
that a holomorphic reparametrization
\be
\U^I \quad \longrightarrow \quad \U'{}^I = f^I \big( \U \big)
\ee
implies
\bea
\F^I \quad & \longrightarrow & \quad \F'{}^I = f^I \big( \F \big) \non \\
\G^I \quad & \longrightarrow & \quad \G'{}^I = \frac{\pa f^I}{\pa \F^J}
\G^J \;.
\eea
Therefore, the whole set of physical scalars
$\{ \F^I, \G^J\}$ parametrizes the
tangent bundle of some K\"ahler manifold.

The action presented in the second line of (\ref{kahler}) is the
so-called chiral--non-minimal nonlinear sigma models \cite{cnm}
which was proposed to describe a supersymmetric low-energy QCD
action.

It is worth pointing out that the computation of the low-energy
hypermultiplet action
(\ref{kahler}) is
rather simple. To determine it, we should in fact evaluate the effective
action for non-vanishing values of the matter chiral superfields
$\F^I$ only.

\sect{Conclusion}
Some years ago it was shown \cite{gio} how to realize the $N=2$ off-shell
matter multiplets with finitely many components fields
(the tensor multiplets \cite{wess,dv,lr2,dpV},
the relaxed hypermultiplet \cite{hst}
and its higher relaxations \cite{ys,gio}, the generalized tensor
or $O(2k)$ multiplets \cite{klt,lr}) in the harmonic superspace approach.
The polar and tropical multiplets,
which are the most interesting, for applications, multiplets in projective
superspace, possess infinitely many auxiliary or purely gauge components.
We have shown in the present paper that these projective multiplets
naturally originate in harmonic superspace as well.

In our opinion, the importance of the projective superspace
approach is that it defines a minimal truncation of unconstrained
analytic superfields, which preserves several fundamental properties
of multiplets and is most suitable for reduction to
$N=1$ superfields. The $q^+$ hypermultiplet cannot be truncated
to a multiplet with finitely many components,
since the charged hypermultiplet
must possess infinitely many auxiliary fields \cite{hsw}.
But it can be decomposed into a sum of two submultiplets, one of which
is just the polar multiplet and the other is purely auxiliary.

The polar and $O(2k)$ multiplets involve, as one of their $N=1$ components,
a remarkable representation of $N=1$ supersymmetry -- the non-minimal
scalar multiplet being described by a complex linear superfield \cite{gs,dg}.
This multiplet has remained for a long time in shadow of the
chiral scalar which is traditionally used to describe supersymmetric
matter. In conventional $N=2$ supersymmetry, the non-minimal scalar
multiplet is seen to be unavoidable. It is worth also remarking that $N=2$
supersymmetry provides us with an explanation of the magic $N=1$
mechanism of generating masses for non-minimal scalars in tandem
with chiral superfields \cite{dg} (see \cite{bk} for a recent review).
Coupling of the polar multiplet to an external $N=2$ vector multiplet
is achived by deforming the polar multiplet constraints via
the covariantization of the $N=2$ covariant derivatives. If we now choose the
background $N=2$ vector multiplet to possess a constant strength,
we result in a massive $N=1$ non-minimal scalar multiplet.

Since projective superspace admits only the $U(1)$ subgroup of
the automorphism group $SU(2)_A$,
it seems to be perfectly suited for formulating
$N=2$ anti-de Sitter supersymmetry as well as for realizing
the $N=2$ higher-superspin massless multiplets \cite{gks} in
a manifestly supersymmetric form. As concerns $N=2$ anti-de Sitter
supersymmetry, it can be most likely realized in harmonic superspace
by choosing a $u$-dependent vacuum solution for a compensator of
$N=2$ supergravity .

\vspace{1cm}

\noindent
{\bf Acknowledgements.}
I am grateful to I.L. Buchbinder,
S.J. Gates, E.A. Ivanov, B.A. Ovrut and E. Sokatchev for valuable
discussions and thankful to M. Ro\v{c}ek for informing me of Refs. \cite{lr}.
It is pleasant to acknowledge a partial support
from RFBR grant, project No 96-02-16017; RFBR-DFG grant,
project No 96-02-00180;
INTAS grant, INTAS-96-0308; grant in the field of fundamental
natural sciences from the Ministry of General and Professional
Education of Russian Federation.

\begin{appendix}
\setcounter{equation}{0}

\section{Tensor fields on the two-sphere}

In this appendix we describe, for completeness, the well-known one-to-one
correspondence between smooth tensor fields on $S^2 = SU(2)/U(1)$ and
smooth scalar functions over $SU(2)$ with definite $U(1)$ charges.
The two-sphere is obtained from $SU(2)$ by factorization
with respect to the equivalence relation
\be
u^{+i} \sim {\rm e}^{{\rm i} \varphi} u^{+i}  \qquad \varphi \in
{\Bbb R} \;.
\label{er}
\ee

We start by introducing two open charts forming an atlas on $SU(2)$
which, upon identificationon (\ref{er}), provides us with
a useful atlas on $S^2$. The north
patch is defined by
\be
 u^{+1} \neq 0
\label{np}
\ee
and here we can represent
\bea
u^{+i} = u^{+1} w^i & \qquad & w^i = (1, u^{+2} / u^{+1}) = (1,w)
 \non \\
u^-_i = \overline{u^{+1}} {\bar w}_i & \qquad & {\bar w}_i = (1, {\bar w})
\qquad |u^{+1}|^2 = (1+ w {\bar w})^{-1}\;.
\eea
The south patch is defined by
\be
 u^{+2} \neq 0
\label{sp}
\ee
and here we have
\bea
u^{+i} = u^{+2} y^i & \qquad & y^i = (u^{+1} / u^{+2},1) = (y,1)
 \non \\
u^-_i = \overline{u^{+2}} {\bar y}_i & \qquad & {\bar y}_i = ({\bar y},1)
\qquad |u^{+2}|^2 = (1+ y {\bar y})^{-1}\;.
\eea
In the overlap of the two charts we have
\be
u^{+i} = \frac{ {\rm e}^{{\rm i} \a} }{ \sqrt{(1+ w {\bar w})} } \;w^i =
\frac{ {\rm e}^{{\rm i} \b} }{ \sqrt{(1+ y {\bar y})} } \;y^i
\ee
where
\be
y = \frac{1}{w} \qquad \quad {\rm e}^{{\rm i} \b} =
\sqrt{\frac{w}{\bar w}} \;{\rm e}^{{\rm i} \a}\;.
\ee
The variables $w$ and $y$ are seen to be local complex coordinates
on $S^2$ considered as the Riemann sphere, $S^2 = {\Bbb C} \cup \{\infty\}$;
the north chart $U_{\rm N} = {\Bbb C}$ is parametrized by $w$
and the south patch $U_{\rm S} = {\Bbb C}\,{}^* \cup \{\infty\}$ is
parametrized by $y$.

Along with $w^i$ and ${\bar w}_i$, we often use their
counterparts with lower (upper) indices
\be
w_i = \ve_{ij} w^j = (-w, 1) \qquad {\bar w}^i = \ve^{ij} {\bar w}_j
= ({\bar w}, -1) \qquad \overline{w_i} = - {\bar w}^i
\ee
and similar for $y_i$ and ${\bar y}^i$.

Let $\F^{(p)}(u)$ be a smooth function on $SU(2)$ with $U(1)$-charge $p$
which we choose, for definiteness, to be non-negative, $p \geq 0$.
Such a function possesses a convergent Fourier series of the form
\be
\F^{(p)}(u) = \sum_{n=0}^{\infty} \F^{(i_1 \cdots i_{n+p} j_1 \cdots j_n)}
u^+_{i_1} \cdots u^+_{i_{n+p}} u^-_{j_1} \cdots u^-_{j_n} \;.
\label{smoothfunction}
\ee
In the north patch we can write
\bea
\F^{(p)}(u) &=& (u^{+1})^p \; \F^{(p)}_{{\rm N}}(w, {\bar w}) \non \\
\F^{(p)}_{{\rm N}}(w, {\bar w}) &=&
\sum_{n=0}^{\infty} \F^{(i_1 \cdots i_{n+p} j_1 \cdots j_n)} \;
\frac{ w_{i_1} \cdots w_{i_{n+p}} {\bar w}_{j_1} \cdots {\bar w}_{j_n} }
{(1 + w {\bar w})^n }\;.
\label{fn}
\eea
In the south patch we have
\bea
\F^{(p)}(u) &=& (u^{+2})^p \; \F^{(p)}_{{\rm S}}(y, {\bar y}) \non \\
\F^{(p)}_{{\rm S}}(y, {\bar y}) &=&
\sum_{n=0}^{\infty} \F^{(i_1 \cdots i_{n+p} j_1 \cdots j_n)} \;
\frac{ y_{i_1} \cdots y_{i_{n+p}} {\bar y}_{j_1} \cdots {\bar y}_{j_n} }
{(1 + y {\bar y})^n }\;.
\eea
Finally, in the overlap of the two charts $\F^{(p)}_{{\rm N}}$ and
$\F^{(p)}_{{\rm S}}$ are simply related to each other
\be
\F^{(p)}_{{\rm S}}(y, {\bar y}) = \frac{1}{w^p}
\; \F^{(p)}_{{\rm N}}(w, {\bar w})\;.
\ee
If we redefine
$$
\hat{\F}^{(p)}_{{\rm N}}(w, {\bar w})
= {\rm e}^{ {\rm i} p \pi/4 } \;\F^{(p)}_{{\rm N}}(w, {\bar w})
\qquad  \check{\F}^{(p)}_{{\rm S}}(y, {\bar y})=
{\rm e}^{ -{\rm i} p \pi/4 }\; \F^{(p)}_{{\rm S}}(y, {\bar y})
$$
the above relation takes the form
\be
\check{\F}^{(p)}_{{\rm S}}(y, {\bar y})
= \left( \frac{\pa y}{\pa w} \right)^{p/2} \;
\hat{\F}^{(p)}_{{\rm N}}(w, {\bar w})
\ee
and thus defines a smooth tensor field on $S^2$.

In accordance with eq. (\ref{breve}), the smile-conjugate of
function (\ref{smoothfunction}) reads
\be
\breve{\F}^{(p)}(u) = (-1)^p \sum_{n=0}^{\infty}
{\bar \F}^{(i_1 \cdots i_{n+p} j_1 \cdots j_n)}
u^+_{i_1} \cdots u^+_{i_{n+p}} u^-_{j_1} \cdots u^-_{j_n} \;.
\label{scsmoothfunction}
\ee
It is easy to check that $\breve{\F}^{(p)}_{\rm S} (y, {\bar y})$
is obtained from $\F^{(p)}_{\rm N} (w, {\bar w})$ by composing
the complex conjugation with replacement $w \rightarrow - {\bar y}$,
\be
\breve{\F}^{(p)}_{\rm S} (y, {\bar y}) =
{\bar \F}^{(p)}_{\rm N} (w, {\bar w})\Big|_{w \longrightarrow - {\bar y}} \;.
\ee
If $p$ is even, in the overlap of the north and south charts we
can represent
\be
\F^{(2k)} (u) = ({\rm i} u^{+1} u^{+2})^k \F^{(p)}_{\rm N-S} (w, {\bar w})\;.
\label{ns}
\ee
Then
\be
\breve{\F}^{(2k)} (u) = ({\rm i} u^{+1} u^{+2})^k
\breve{\F}^{(2k)}_{\rm N-S} (w, {\bar w})
\ee
where $\breve{\F}^{(2k)}_{\rm N-S} (w, {\bar w})$ is obtained from
$\F^{(2k)}_{\rm N-S} (w, {\bar w})$ by composing
the complex conjugation with replacement $w \rightarrow -
\frac{1}{{\bar w}}$,
\be
\breve{\F}^{(2k)}_{\rm N-S} (w, {\bar w}) =
{\bar \F}^{(2k)}_{\rm N-S} (w, {\bar w})\Big|_{w \longrightarrow -
\frac{1}{\bar w}} \;.
\label{defreal}
\ee
From here we recover the projective superspace conjugation
(\ref{pconjugation}).

\setcounter{equation}{0}

\section{$O(2k)$ multiplet in harmonic superspace}

The $O(2k)$ multiplet is described in harmonic
superspace \cite{gio}
by an analytic real superfield $\O^{(2k)} (z, u)$,
\be
D^+_\a \O^{(2k)} = {\bar D}^+_\ad \O^{(2k)} = 0 \qquad \qquad
\breve{\O}^{(2k)} = \O^{(2k)}\;,
\label{o2kconstr1}
\ee
which, in addition, is constrained by
\be
D^{++} \O^{(2k)} = 0\;.
\label{o2kconstr2}
\ee
This constraint along with the analyticity conditions imply
\be
\O^{(2k)} (z,u) = \O^{i_1 \cdots i_{2k}} (z) u^+_{i_1} \cdots u^+_{i_{2k}}
\qquad
D_\a^{(j} \O^{i_1 \cdots i_{2k})}
= {\bar D}_\ad^{(j} \O^{i_1 \cdots i_{2k})} = 0\;.
\ee
In the north chart we can represent
\be
\O^{(2k)} (z,u) = \frac{1}{(2k-2)!}
\big({\rm i}\, u^{+1} u^{+2}\big)^k \O^{[2k]}(z,w)
\ee
where $\O^{[2k]}$ is given by eq. (\ref{ro(k)}). The action reads
\be
S = \hf \,(4k-3)!\,
\int {\rm d}u \,{\rm d}\z^{(-4)} \, \big[ (u^- s) (u^-{\bar s}) \big]^{2k-2}
\left(\O^{(2k)}\right)^2\;.
\ee
Here $s$ and ${\bar s}$ are the constant isospinors (\ref{punctures})
defining the south and north poles.
In spite of the fact that the constraints (\ref{o2kconstr1}) and
(\ref{o2kconstr2}) are $SU(2)_A$ covariant, for $k > 1$ the action
is invariant only with respect to the $U(1)$ subgroup of $SU(2)_A$.

\end{appendix}

\end{document}